\documentclass[aps,onecolumn,groupedaddress,showpacs,notitlepage,nofootinbib]{revtex4-1}

\usepackage{graphicx}
\usepackage{tikz}
\usepackage{tikz-feynman}
\usepackage{amssymb,mathrsfs,epsfig,color, url}
\usepackage{amsmath}
\usepackage{amsfonts}
\usepackage{booktabs}
\usepackage{colortbl}
\usepackage{hyperref}
\usepackage{slashed}
\usepackage{fancyhdr}
\usepackage[ddmmyyyy,hhmmss]{datetime}
\usepackage{filecontents}
\usepackage{float}
\usepackage{caption}
\usepackage{subcaption}
\usepackage{tabu}
\usepackage[justification=centering]{caption}
\usetikzlibrary{external}
\newcommand{\GeV}{~\mathrm{GeV}}
\newcommand{\TeV}{~\mathrm{TeV}}

\newcommand{\RD}{R_{D^{(*)}}}
\newcommand{\RK}{R_{K^{(*)}}}

\numberwithin{equation}{section}

\newcommand{\be}{\begin{equation}}
\newcommand{\ee}{\end{equation}}
\newcommand{\bea}{\begin{eqnarray}}
\newcommand{\eea}{\end{eqnarray}}

\newcommand{\overbar}[1]{\mkern 1.5mu\overline{\mkern-1.5mu#1\mkern-1.5mu}\mkern 1.5mu}

\def\LQ{\textit{LQ}}

\begin{document}
%\today

\title{ Probing leptoquark chirality via top polarization at the Colliders }

%\author{Yu Gao$^{1}$}
%\email{gaoyu@ihep.ac.cn}
\author{Joydeep Roy}
\email{jdroy@itp.ac.cn}

\affiliation{CAS Key Laboratory of Theoretical Physics, Institute of Theoretical Physics,\\
Chinese Academy of Sciences, Beijing 100190, China}

\begin{abstract}
Anomalies in recent LHCb, Belle and Babar measurements of $R_{D^{(*)}}$, and $R_{K^{(*)}}$ in $B$ decays may indicate the new physics beyond the Standard Model (SM). The leptoquarks ($LQ$) that couple to the $3^{\mathrm{rd}}$ generation quarks and leptons have been proposed as a viable new physics (NP) explanation. Such left-handed $LQ$s can couple to both bottom and top quarks. Since top particles decay before the hadronization, it is possible to reconstruct chirality of boosted top quarks and consequently the chirality of top coupling to the $LQ$s. We perform analysis on the top quark's chirality in the pair-production channel of the $LQ$, which can be purely left-handed in comparison to unpolarized $t\bar{t}$ SM background.  We study the prospects of distinguishing the chirality of a potential $LQ$  signal for the high luminosity run of the LHC and other future colliders.
\end{abstract}

\maketitle

%%%%%%%%%%%%%%%%%%%%%%%%%%%%%%%%%%%%%%%%%%%%%%%%%%%%%%%%%%%%

\section{Introduction}

Leptoquarks generally exist in grand unification models that mediate the transition between lepton and quark fermions.   The Standard Model (SM) is gauge anomaly free due to the cancellation between the lepton and quark contributions. Although the  baryon number ($B$) and the lepton number ($L$) are conserved in the Standard Model, both $B, L$ violation generally occur in the extended symmetry structures in the grand unification models~\cite{bib:GUTS} like the $SU$(5), $SU$(10), $E_6$, etc. while a $B-L$ number conservation can still be preserved~\cite{Heeck:2014zfa}. Explicit $B$ or $L$ violating leptoquark couplings can lead to fast proton decay, and such leptoquarks need to be very heavy to avoid stringent experimental constraints~\cite{Dorsner:2016wpm}. Alternatively the $B$ and $L$ violating terms can also be forbidden by imposing discrete $Z_3$ symmetry on the SM lepton and quark fields~\cite{Arnold:2013cva}. On the other hand, $B,L$ conserving $\LQ$s which can avoid proton decay and are light enough, are much less constrained and they lead to rich phenomenology at current collider and cosmic ray searches (see Ref.~\cite{Dorsner:2016wpm} for recent reviews and references therein).

The flavor structure of $\LQ$ couplings can lead to many interesting new physics signals. Leptoquarks in general can couple to multiple SM fermion generations/flavors. A leptoquark that couples to different lepton or quark generations contributes to loop-level flavor-changing neutral currents~\cite{Fajfer:2008tm} and is tightly constrained for the first two generations~\cite{Leurer:1993em,Sahoo:2015wya}. Even for flavor-diagonal leptoquark couplings, flavor non-universal couplings contribute to the violation of Lepton Flavor Universality (LFU) and yield large correction to the (semi)leptonic branching ratios in heavy-flavored meson decays~\cite{Sahoo:2015pzk}. Recently $\LQ$s have received much attention as possible explanation~\cite{bib:BAnoLQ} to the LFU anomaly in $B$ meson's decays that has been reported by BaBar~\cite{Lees:2013uzd}, LHCb~\cite{Aaij:2015yra} and Belle~\cite{Abdesselam:2016cgx} experiments, in which the branching ratios
$\RD = \mathrm{Br}(\bar{B} \rightarrow D^{(*)}\tau^-\bar{\nu}_{\tau})/\mathrm{Br}(\bar{B} \rightarrow D^{(*)}l^-\bar{\nu}_{l})$ and $\RK = \mathrm{Br}(B^0 \rightarrow K^{(*0)}\mu^+\mu^-)/\mathrm{Br}(B^0 \rightarrow K^{(*0)}e^+e^-)$ are at $3.9\sigma$~\cite{Aaij:2017deq} and $2.1-2.5\sigma$~\cite{Aaij:2015yra} deviation respectively from their SM predictions. $\LQ$s that couple to the third generation of quarks and leptons can generate the $b$ quark - lepton interaction to explain these anomalies and stay consistent with the current experimental bounds \cite{Fajfer:2015ycq}-\cite{Sahoo:2016pet}.

Leptoquarks can couple differently to the left and right handed (chiral) fermion currents. A coupling to certain lepton and quark chirality can be the indicator of the leptoquark's identity. At the LHC, the chiral property of $\LQ$ can be probed if the $\LQ$s, after their potential pair production from QCD interactions, decay to heavy-flavor fermions whose polarization can be reconstructed from the subsequent decays. While the $\mu,\tau$ chirality is difficult to obtain at the LHC due to their long lifetimes and partially invisible decay products, the $t$ quark is an optimal candidate and its chirality can be statistically identified in both full hadronic and semileptonic $t$ decay channels~\cite{Allahverdi:2015mha}. In contrast, other heavy-flavor quarks hadronize before decaying and lose their chirality information. In this work, we focus on the types of $\LQ$s that couple to the $t$ quark and study the prospects of probing the leptoquark's chiral nature of the coupling by reconstructing the top chirality from leptoquark decays at the LHC or future high-luminosity colliders.

We briefly discuss the chiral and spin category of leptoquarks in Section~\ref{Sec:LQmodel} and identify the candidates that are relevant for the top-quark chirality search at the LHC. Section~\ref{Sec:Topchirality} studies the potential collider signatures and strategies of distinguishing the signal from a dominant SM $t\bar{t}$ background. Detailed simulation analysis is represented in Sec~\ref{Sec:Analysis}. We make a HL-LHC and future collider sensitivity study and show our results in Section~\ref{Sec:Results} and then conclude in Section~\ref{Sec:conclusion}.

%%%%%%%%%%%%%%%%%%%%%%%%%%%%%%%%%%%%%%%%%%%%%%%%%%%%%%%%%%%%

\section{The Leptoquark Model}
\label{Sec:LQmodel}

%Here we discuss the details of the leptoquark model we used.

\subsection{Vector Leptoquark, $U_3=(\mathbf{3},\mathbf{3},2/3)$}

Leptoquarks are fields that can simultaneously couple to a lepton and a quark field. Depending on spin there are two types of $\LQ$s, spin-zero (scalar) and spin-one (vector). Since it can couple to both lepton and quark, we can also classify them according to the standard model (SM) representations. Also under the SM gauge group there exist six scalar and six vector $\LQ$ multiplets \cite{Dorsner:2016wpm} whose representations, symbols and the chiralities are shown in Table \ref{tab:LQs}. The $\LQ$s which differ only in the hypercharge, are denoted by a tilde or bar over their symbol. Since the right chiral neutrino's presence is till now a questionable issue, an additional bar is put on those types of \textit{LQ}s which contain them.

\begin{table}[H]
\centering
\begin{tabular}{|c|c|c|}
\hline
$(SU(3)_C,SU(2)_L,U(1)_Y)$ &  Symbol & Type  \\
\hline 
$(\overline{\mathbf{3}},\mathbf{3},1/3)$  & $S_3$ & $LL$\,$(S^L_1)$  \\
$(\mathbf{3},\mathbf{2},7/6)$  & $R_2$ & $RL$\,$(S^L_{1/2})$, $LR$\,$(S^R_{1/2})$  \\
$(\mathbf{3},\mathbf{2},1/6)$  & $\tilde{R}_2$ & $RL$\,$(\tilde{S}^L_{1/2})$, $\overline{LR}$\,$(\tilde{S}^{\overline{L}}_{1/2})$  \\
$(\overline{\mathbf{3}},\mathbf{1},4/3)$  & $\tilde{S}_1$ & $RR$\,$(\tilde{S}^R_{0})$  \\
$(\overline{\mathbf{3}},\mathbf{1},1/3)$  & $S_1$ & $LL$\,$(S^L_0)$, $RR$\,$(S^R_0)$, $\overline{RR}$\,$(S^{\overline{R}}_0)$  \\
$(\overline{\mathbf{3}},\mathbf{1},-2/3)$  & $\bar{S}_1$ & $\overline{RR}$\,$(\bar{S}^{\overline{R}}_0)$  \\
\hline
\hline
$(\mathbf{3},\mathbf{3},2/3)$  & $U_3$ & $LL$\,$(V^L_1)$ \\
$(\overline{\mathbf{3}},\mathbf{2},5/6)$  & $V_2$ & $RL$\,$(V^L_{1/2})$, $LR$\,$(V^R_{1/2})$  \\
$(\overline{\mathbf{3}},\mathbf{2},-1/6)$  & $\tilde{V}_2$ & $RL$\,$(\tilde{V}^L_{1/2})$, $\overline{LR}$\,$(\tilde{V}^{\overline{R}}_{1/2})$  \\
$(\mathbf{3},\mathbf{1},5/3)$  & $\tilde{U}_1$ & $RR$\,$(\tilde{V}^R_0)$  \\
$(\mathbf{3},\mathbf{1},2/3)$  & $U_1$ & $LL$\,$(V^L_0)$, $RR$\,$(V^R_0)$, $\overline{RR}$\,$(V^{\overline{R}}_0)$  \\
$(\mathbf{3},\mathbf{1},-1/3)$  & $\bar{U}_1$ & $\overline{RR}$\,$(\bar{V}^{\overline{R}}_0)$  \\
\hline 
\end{tabular}
\caption{\label{tab:LQs} List of scalar and vector $\LQ$s.}
\end{table} 

Out of these twelve $\LQ$ multiplet, it is shown \cite{Calibbi:2017qbu, Alok:2017sui} that a scalar $\LQ$ isotriplet with hypercharge $(Y = 1/3)$, a vector $\LQ$ isotriplet with hypercharge $(Y = -2/3)$ and a vector $\LQ$ isosinglet with hypercharge $(Y = -2/3)$ can contribute to the Flavor-changing-neutral-currents (FCNCs) and thus to recently observed $B$-physics anomalies. Following the nomenclature of Ref.\cite{Dorsner:2016wpm} they are called $S_3, U_1\ \textrm{and}\ U_3$ respectively. Also depending on the chirality of both quark and lepton the chirality of the quark-lepton-$\LQ$ operator is determined. Therefore following the same nomenclature we find that $S_1$ and $U_3$ are of purely left-handed in nature and since we are interested in a vector $\LQ$ with pure left-handed couplings, we chose $U_3$ with the SM gauge group representation $(\mathbf{3},\mathbf{3}, 2/3)$ as our study sample. Some recent analysis \cite{Buttazzo:2017ixm,Angelescu:2018tyl} have also shown that $U_1$ without the right-hand couplings can also contribute to the recent $B$-anomalies and this study therefore can also be relevant for those leptoquarks. Similarly, if one is interested in probing the chirality of a pure left-handed scalar $\LQ$, the ideal candidate would be $S_3$.

The corresponding term that goes into the Lagrangian is given by \cite{Dorsner:2016wpm}

\be \label{eq:main_U_3}
\mathcal{L}_{\LQ} = g_{\LQ} \bar{Q}^a_L \gamma^{\mu} L_L (V)^a_{\mu} + \textrm{h.c.}
\ee 
where $Q^a_L (L_L)$ is the left-handed quark (lepton) doublet of SM, subscript $L$ stands for the Left-handed projection operator $P_L = (1-\gamma_5)/2 $ and $g_{\LQ}$ is the corresponding coupling of the leptoquark with the SM particles.

At hadronic collision the $\LQ$s can be produced either singly or in pairs. The single production is largely model dependent due to unknown Yukawa couplings. The pair production of leptoquarks happen via QCD interactions because these vector $\LQ$s are colour triplet  and the interactions depend only on their spin \cite{Rizzo:1996ry}.

The relevant kinetic and mass terms which give the vector leptoquark-gluon interactions include
\be \label{eq:LQQCD}
\mathcal{L}_{VLQ}^{QCD} = -\frac{1}{2}F^{\dagger}_{\mu\nu}F^{\mu\nu} + M_V^2V^{\dagger}_{\mu}V^{\mu} - ig_s\kappa V^{\dagger}_{\mu}G^{\mu\nu}V_{\nu},
\ee
where $G^{\mu\nu}$ is the usual gluon field strength tensor, $V_{\mu}$ is the vector leptoquark field, $F^{\mu\nu} = D_{\mu}V_{\nu} - D_{\nu}V_{\mu} $ is the field strength tensor with $SU(3)$ covariant derivative $D_{\mu} = \partial_{\mu} + ig_sT^aG^a_{\mu}$, $G^a_{\mu}$ is the gluon field and $T^a$ is the $SU(3)$ genarator. $\kappa$ is known as `anomalous chromomagnetic moment' which is usually taken to be 1 for numerical calculations \cite{Rizzo:1996ry}. This is known as the minimum coupling case, though other possibilities are also considered in literature. The advantage of taking $\kappa = 1$ is that depending on the \textit{LQ} mass, the production cross-section of a vector \textit{LQ} is increased to $5-20$ times larger than that of the scalar one. 

\iffalse
%\centering
\begin{subfigure}{0.1\textwidth}
\feynmandiagram [horizontal=a to b] {
  i1[particle=\(g\)] -- [gluon] a -- [gluon] i2[particle=\( g\)],
  a -- [gluon] b,
  f1 [particle=\(\overbar{\LQ}\)] -- [scalar] b -- [scalar] f2 [particle=\(\LQ\)],
};
\end{subfigure}
\hspace{2.5 cm}
\begin{subfigure}{0.1\textwidth}
\feynmandiagram [horizontal=i1 to f2] {
  i1[particle=\(g\)] -- [gluon] a -- [gluon] i2[particle=\( g\)],
  f1 [particle=\(\overbar{\LQ}\)] -- [scalar] a -- [scalar] f2 [particle=\(\LQ\)],
};
\end{subfigure}
\hspace{2.5 cm}
\begin{subfigure}{0.1\textwidth}
\feynmandiagram [vertical=a to b] {
  i1[particle=\(g\)] -- [gluon] b -- [scalar] f1[particle=\(\LQ\)],
  a -- [scalar,edge label=\(\LQ\)] b,
  i2 [particle=\( g\)] -- [gluon] a -- [scalar] f2 [particle=\(\overbar{\LQ}\)],
};
\end{subfigure}
\hspace{1.5 cm}
\begin{subfigure}{0.1\textwidth}
\feynmandiagram [horizontal=a to b] {
  i1[particle=\(q\)] -- [fermion] a -- [fermion] i2[particle=\(\bar q\)],
  a -- [gluon] b,
  f1 [particle=\(\overbar{\LQ}\)] -- [scalar] b -- [scalar] f2 [particle=\(\LQ\)],
};
\end{subfigure}
\fi

\begin{figure}[H]
\centering
\hspace*{\fill}
  \includegraphics[width=17cm]{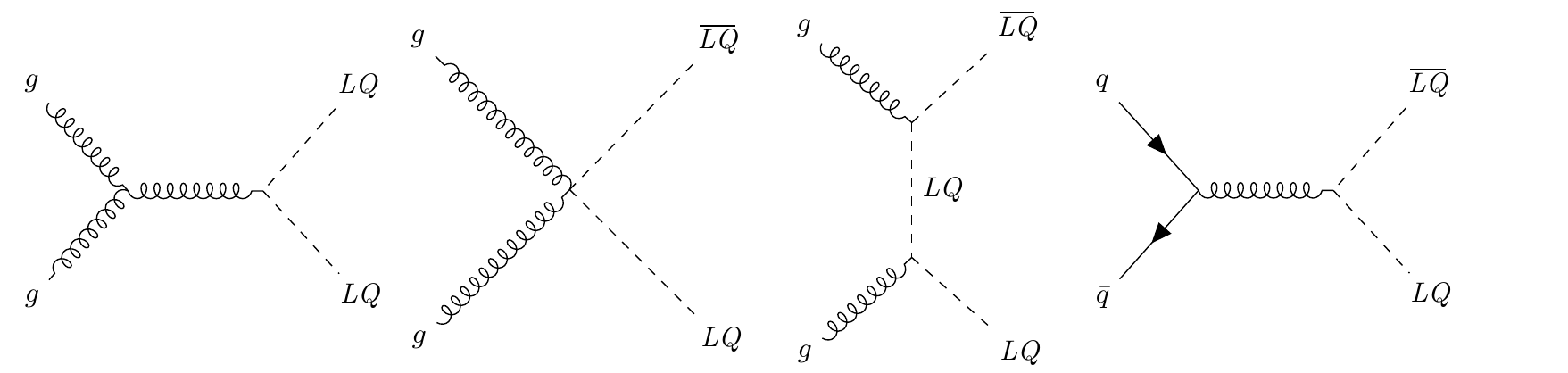}
\caption{LQ pair production from strong interactions. The dominant contributions to the $\LQ$ production come from the trilinear $gVV$ and quartic $ggVV$ couplings}
\label{Fig:LQ from QCD}
\end{figure}

Fig.\ref{Fig:LQ from QCD} shows the dominant leading-order (LO) diagrams for leptoquark pair-production from all possible strong interactions including the quark-quark and gluon fusion.
In order to calculate the production cross-section of pair of vector leptoquarks ($VLQ$) from gluon fusion, $gg \rightarrow VV$,  we need to determine both the trilinear $gVV$ and quartic $ggVV$ couplings. Usually these couplings are fixed by extended gauge invariance.

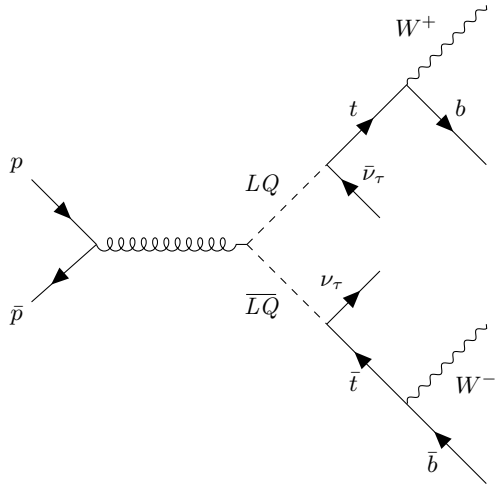
\begin{figure}[H]
\centering
\begin{tikzpicture}
\begin{feynman}
\vertex (a1) {\(p\)};
\vertex [below=2.0cm of a1] (a2) {\(\bar p\)};
\vertex [below right=1.5cm of a1] (b);
\vertex [right=2.0cm of b] (c);
\vertex [above right=of c] (d1);
\vertex [below right=of c] (d2);
\vertex [above right=of d1] (d11);
\vertex [below right=1.0cm of d1] (f3);
\vertex [above right=1.0cm of d2] (f4);
\vertex [below right=of d2] (d22);
\vertex [above right=of d11] (f1);
\vertex [below right=of d11] (f2);
\vertex [above right=of d22] (f5);
\vertex [below right=of d22] (f6);

\diagram* {
 (a1) -- [fermion] (b),
 (a2) -- [anti fermion] (b),
 (b) -- [gluon] (c),
 (c) -- [scalar,edge label=\(\LQ\)] (d1),
 (c) -- [scalar,edge label'=\(\overbar{\LQ}\)] (d2),
 (d1) -- [fermion, edge label=\(t\)] (d11),
  (d1) -- [anti fermion, edge label=\(\bar \nu_{\tau}\)] (f3),
  (d2) -- [anti fermion, edge label'=\(\bar t\)] (d22),
  (d2) -- [fermion, edge label=\(\nu_{\tau}\)] (f4),
  (d11) -- [boson, edge label=\(W^{+}\)] (f1),
  (d11) -- [fermion, edge label=\(b\)] (f2),
  (d22) -- [boson, edge label'=\(W^{-}\)] (f5),
  (d22) -- [anti fermion, edge label'=\(\bar b\)] (f6),
};
\end{feynman}
\end{tikzpicture}
\caption{Cascade decay of a vector $\LQ$. }
\label{Fig:Vector LQ Decay}
\end{figure}

Fig.\ref{Fig:Vector LQ Decay} shows a conventional leptoquark decay chain where it couples to the top quark. In the collider the signature of both the single and pair-production of $\LQ$s are usually similar and they are given by \cite{Belyaev:2005ew}
\begin{enumerate}
\item $2l + jets$ for $\LQ$s decaying into a lepton and a quark;
\item $l + jets + \slashed{E_T} $ for one $\LQ$ decaying into a lepton and a quark and the other into a neutrino and a quark;
\item $  jets + \slashed{E_T} $ both $\LQ$s decaying into a neutrino and a quark.
\end{enumerate}
Clearly, according to Fig.\ref{Fig:Vector LQ Decay} we are considering the $3^{rd}$ signature. Similar channel has been explored in Ref. \cite{Vignaroli:2018lpq}. The possible backgrounds for the pair-production of the $\LQ$s may be QCD or the gauge boson production with jets, but the dominant contribution comes from the $t\bar{t}$ production.

\subsection{The Feynrules model file}

We used \texttt{FEYNRULES}\cite{Christensen:2008py} package to create our leptoquark model. Following the above discussion to generate this model we just need to add the extra Lagrangian terms to the usual Standard Model (SM) Lagrangian. Therefore including the equations \ref{eq:main_U_3} and \ref{eq:LQQCD} the final Lagrangian for this model becomes
\be \label{eq:Final Lagrangian}
\mathcal{L}_{\LQ}^{Final} = \mathcal{L}_{SM} +  \mathcal{L}_{\LQ} + \mathcal{L}_{VLQ}^{QCD}.
\ee
The new relevant parameters that need to be defined within the \texttt{FEYNRULES} framework were the $\LQ$ field itself and its coupling. To give mass to our \textit{LQ} we were trying to be consistent with the recent exclusion limits for the vector \textit{LQ}s \cite{Sirunyan:2018kzh}. Also we were aware about the fact that for a heavy \textit{LQ} the cross-section is significantly low at the LHC. Therefore we chose to give a mass of $1.5 \TeV$ to this $\LQ$. Since the NP explanation to the $B$-physics anomalies consider mainly the coupling of $\LQ$ to $3^\mathrm{rd}$ generation fermions, we took $g_{\LQ}$ to be unity.  To use for our analysis an Universal Feynrules Output (UFO) \cite{Degrande:2011ua} was generated using these parameters and the Lagrangian \ref{eq:Final Lagrangian}. As shown in Fig \ref{Fig:Vector LQ Decay}, the decay of $W$s produce lots of jets in the collider as well as significant amount of missing energy $\slashed{E}_T$. To reconstruct a two-\textit{top} systems which are eventually the decay product of $\LQ$s, we chose one $b$-jet and two jets for each top particle. Therefore we finally took two \textit{b}-jets and four non-\textit{b}-jets to constitute a reconstructed event.

%%%%%%%%%%%%%%%%%%%%%%%%%%%%%%%%%%%%%%%%%%%%%%%%%%%%%%%%%%%%

\section{Chirality of top quark as a discriminator}\label{Sec:Topchirality}

Prime motivation of this project was to use the idea of Ref.\cite{Allahverdi:2015mha} where the chirality of top quark was used as a discriminator between the signal of new interaction and the possible background. We know based on the possible polarizations for the $W$, the preferred spin configuration decay channel for the top quark is a longitudinal $W$ boson $(\sim 70\%)$ and a bottom $(b)$ quark, with the top and bottom spin aligned in the same direction in the center-of-mass frame. Now, if the top is left-handed, the momentum of $b$ quark would be parallel to its spin and thus to the Lorentz boost  and since $W$ only couples to the left-handed current the bottom quark would be more energetic in the lab frame. This will create a slanted spectra with the observable, known as $b$ energy ratio and defined as $E(b)/E(t)$. On the contrary due to the equal presence of particles of both chirality the top pair production generates an unpolarized flat spectra. 
Thus with this observable the top polarization can be determined and distinguished easily that we shall show in the following section.  

%%%%%%%%%%%%%%%%%%%%%%%%%%%%%%%%%%%%%%%%%%%%%%%%%%%%%%%%%%%%

\section{Simulation Analysis}
\label{Sec:Analysis}

Our goal is to distinguish the pair-production channel of leptoquarks from the possible backgrounds that is present at the collider. Therefore for our analysis, we separately produced signal and background events, applied several cuts and compared them afterwords. For both, signal and background, we simulated 500k events. We used \texttt{MADGRAPH v3.0}\cite{Alwall:2014hca} for the event genaration, \texttt{PYTHIA8} \cite{Sjostrand:2014zea} for the parton showering and hadronization and \texttt{DELPHES 3.4.1} \cite{deFavereau:2013fsa} for the detector simulation. For this detector level study we used \texttt{.LHCO} files generated in the \texttt{MADGRAPH v3.0}.

The process used for the $\LQ$ production was $pp > LQ \overbar{LQ}$ while for the background $pp > t\bar{t}$ process was used. A cross-section which is within $20\%$ of the recent top quark pair production cross section measurement reported by CMS \cite{Khachatryan:2015uqb} was obtained and thus we used the CMS value for our analysis purpose. 

\subsection{Cuts used}
\label{Subsec:Cuts}

Several cuts were used to separate the signal events from the background. Since we decayed our vector $\LQ$ to top particles via such processes:
 $pp > LQ \ \overbar{LQ}, LQ > t \ \bar{\nu_{\tau}},\ t >  j j\ b,\ \overbar{LQ} >  \bar{t}\ \nu_{\tau},\ \bar{t} > j j\ \bar{b} $, as shown in the Fig.\ref{Fig:Vector LQ Decay}, the basic idea was to reconstruct the system with two top particles with some specific energy cuts relevant for the analysis. Top being decayed hadronically in this process, it is fully possible to reconstruct the top energy as also done in LHC. Upon successful \textit{b}-tagging, the top energy is given by $E(b)+E(j_1)+E(j_2)$, where $j_1$ and $j_2$ are the two non-\textit{b} jets produced after top decay.
The systematics of our analysis are the following:
 At first, for each event of the \texttt{.LHCO} file we extracted all jets,  means both \textit{b}-tagged jets and non \textit{b}-tag jets. Then we discarded the events which contain less than four non \textit{b}-jets and less than two \textit{b}-jets in it. Since we want to reconstruct 2-top systems, we used the cut on the mass of $W$ particle such that $|m_W - 80.385| \leq 20 \GeV$ and a similar cut on the mass of top particle, $|m_{\rm top} - 173.3| \leq 20 \GeV$. Therefore we took only those events where the reconstructed mass of the top quarks lie within the range $153 \GeV \leq |m_{\rm top}| \leq 193 \GeV $. Now, since there are at least two $b$-jets and four non-\textit{b} jets (total six jets) in each selected event, the minimum number of possible combinations of getting two-top systems would be at least six and since we choose two of them there is a possibility that both of those two-top systems might contain the same non \textit{b}-jet. This is definitely not the desired situation and therefore a further filtration is needed so that this don't happen. Obviously, when there are more than six jets present in an event there would be more possible combinations and a more careful filtering process can be applied, but that is not necessary for our purpose because we are only concerned about the events which contain two reconstructed top masses within the specified range. Now, we are left with the events from which a top pair can be reconstructed. Finally, we took all the jets of each reconstructed top and calculated the energy of them ($E_t$). We used an energy cut of $250 \GeV$ to filter all the elements of the list to get the final list of energies. The reason for putting $250 \GeV$ energy cut is to have a significant amount of signal events compared to the background with a clear distinction between the spectra. We tried some lower cuts such as $100 \GeV \mathrm{or} ~150 \GeV$, as well as some upper cut like $300 \GeV$ to find that for those lower cuts there were significantly large background events and for upper cuts there were considerably smaller signal events. This is quite evident from the energy-cut efficiency for both the signal and the background presented in the Table~\ref{Tab:Cut-efficiency flow table}, as we are having a comparable percentage of events after the energy cut.
We also calculated the energy of \textit{b}-jets ($E_b$) of each element to calculate the ratio ($E_b/E_t$). Finally to remove the SM background maximally, we did our  MET (Missing Transverse Energy ($\slashed{E}_T$))cut analysis of $150 \GeV $.

For the background we used $p p \rightarrow t \bar{t}$ process as this has the largest possible contribution and exactly same cuts were applied for this analysis also.

\subsection{Future collider analysis}

We know after the LHC runs are over, there are proposals for new and more energetic colliders such as Future Circular Collider (FCC-hh) at CERN \cite{bib:FCC} with $\sqrt{s} = 100~\TeV$ and Super Proton Proton Collider (SppC) in China \cite{Su:2015sfa}. For our purpose, we also performed an analysis for the FCC energy range. For this analysis we used the inbuilt \texttt{FCC} card of \texttt{DELPHES} and ran the simulation as usual.

There are some significant differences between the \texttt{FCC} Delphes card and the usual \texttt{LHC} \texttt{delphes.card.dat}. Some relevant ones for our purposes can be mentioned as follows: $(i)$ the \textsf{b-tagging} section of the \texttt{FCC} card has been modified quite significantly with extra informations, $(ii)$ \textsf{Jet Flavor Association} section parameters such as \textsf{PartonPTMin} and \textsf{PartonEtaMax} are changed from 1.0 and 2.5 respectively in original \texttt{delphes.card.dat} to 5.0 and 6.0 respectively in \texttt{FCC} card.$(iii)$ Similarly, \textsf{Jet Finder} section parameters such as \textsf{ParameterR} and \textsf{JetPTMin} are changed from 0.5 and 20.0 respectively in \texttt{LHC} card to 0.4 and 30.0 respectively in \texttt{FCC} card. $(iv)$ The parameters \textsf{ParameterR} and \textsf{JetPTMin} are changed from 0.5 and 20.0 respectively in \texttt{LHC} card to 0.4 and 5.0 respectively in \texttt{FCC} card, etc.

%%%%%%%%%%%%%%%%%%%%%%%%%%%%%%%%%%%%%%%%%%%%%%%%%%%%%%%%%%%

\section{Results}
\label{Sec:Results}
\subsection{Signal and background spectrum analysis}

Table \ref{Tab:Cut-efficiency flow table} shows the efficiency of the signal events as we keep imposing several cuts as explained in Sec.\ref{Subsec:Cuts}. For LHC, after we filter our events having minimum four non-\textit{b}-jets and two \textit{b} jets, $63\%$ of those events fall within $20 \GeV$ from the \textit{W} mass. Out of those events $85\%$ are within $20 \GeV$ from the top mass.  $24\%$ of these events have two top-system among them. Next, when we apply the energy-cut of $250 \GeV$, $32\%$ of previously remaining events pass through that cut. When we multiply all these efficiencies we finally get $4\%$ of the total sample events that satisfy all the cuts. From the results of FCC analysis we find that the efficiency flow is quite similar to that of the LHC.

\begin{table}[h!]
\centering
\resizebox{\textwidth}{!}
{\begin{tabular}{ |c|c|c|c|c|c|c| } 
 \hline
Collider &  $\sigma\times \mathrm{BR}$ (pb)  & W-Cut Eff. & Top-Cut Eff. & 2-Top system Eff. & Energy-Cut Eff. & Final Eff. \\
 \hline 
 LHC $(\sqrt{s} = 13~\TeV)$ & $2.6\times 10^{-4}$ (Sig) & 0.63  & 0.85 &  0.24 &0.32 & 0.041\\ 
 \cline{2-7}
  & 746 (Bkg) & 0.87  & 0.85 &  0.17 &0.36 & 0.047\\
  \hline 
 & 4.8 (Sig) & 0.87 & 0.84 & 0.18 & 0.36 & 0.046\\
  \cline{2-7}
 FCC $(\sqrt{s} = 100~\TeV)$ & $30\times 10^3$ (Bkg) & 0.62 & 0.94 & 0.57 & 0.34 & 0.113\\
 \hline
\end{tabular}}
\caption{Cut-efficiency flow table for \textit{LQ} decay before MET cut at  different colliders}
\label{Tab:Cut-efficiency flow table}
\end{table}

\begin{figure}[H]
\centering
\hspace*{\fill}
\begin{subfigure}{.45\textwidth}
  \centering
  \includegraphics[width=6cm]{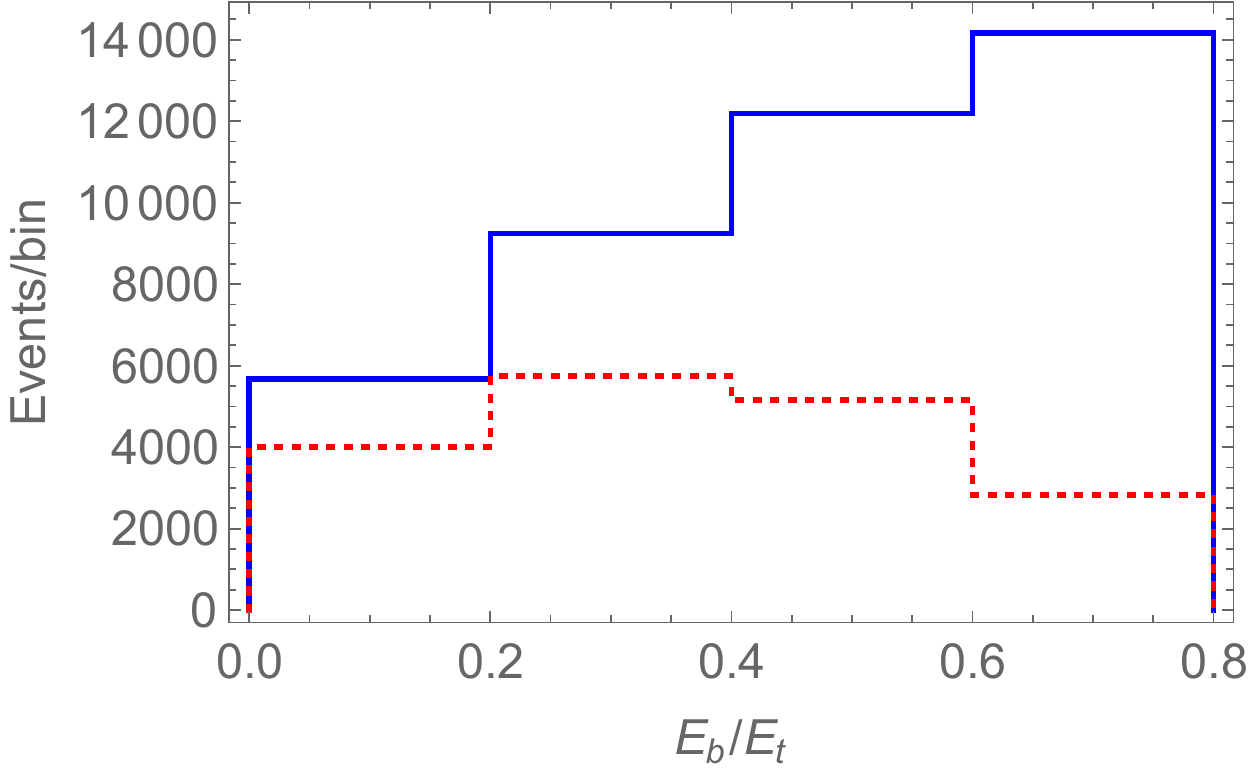}
  \caption{ Comparison of shapes of signal and background for LHC}
  \label{fig:LHCnoMET}
\end{subfigure}%
\hfill
\begin{subfigure}{.45\textwidth}
  \centering
  \includegraphics[width=6cm]{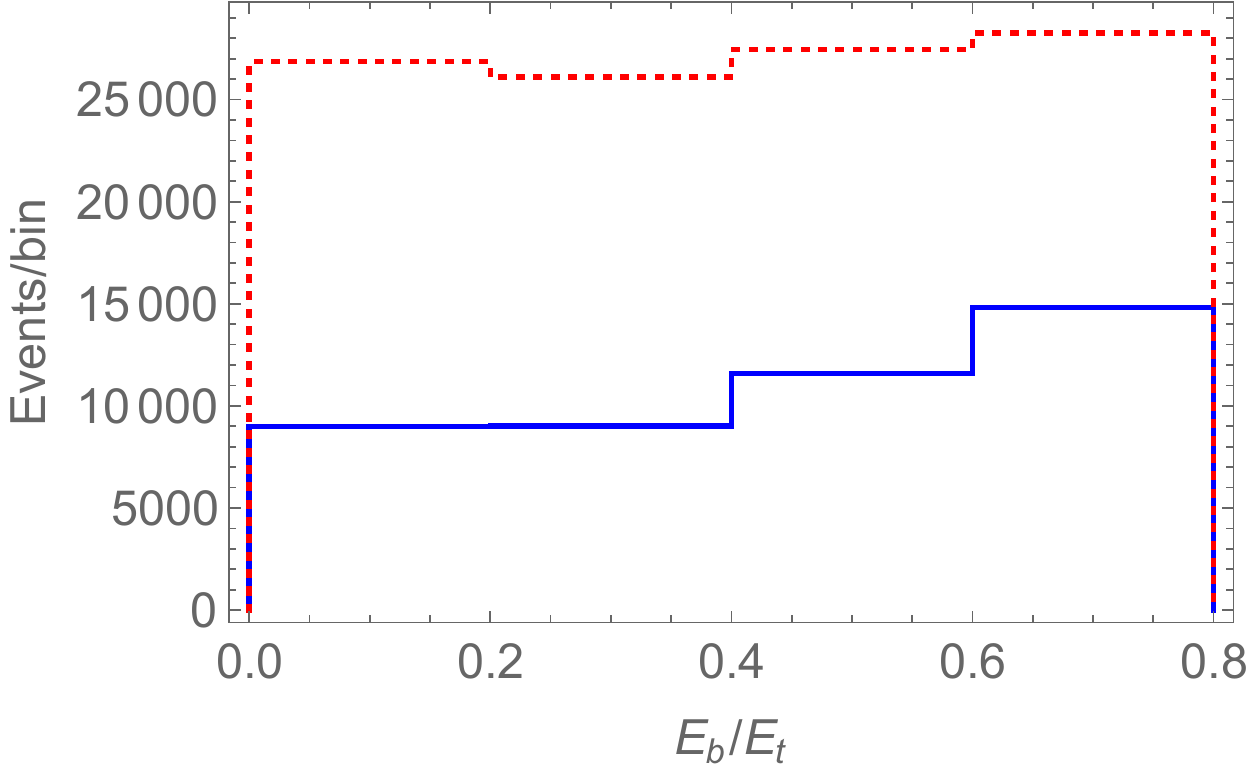}
  \caption{Comparison of shapes of signal and background for FCC}
  \label{fig:FCCnoMET}
\end{subfigure}
\hspace*{\fill}
\caption{ Comparison of shapes of signal and background production cross-section for $1.5 \TeV$ left-handed leptoquark. The after-cut event analysis is done for 4 bins. Blue solid curve represents the $LQ$ signal and the red dotted curve represents the background. Different kinematic cuts (without MET cut) used to produce these plots as described in Section \ref{Sec:Analysis}. }
\label{Fig:LHC-FCC spectrum}
\end{figure}

Fig.\ref{Fig:LHC-FCC spectrum} represents the spectral comparison of the signal and the background before putting the MET cut. Blue solid curve represents the $LQ$ signal and the red dotted curve represents the background. A separation between the chirality states (left-handed for the signal (\textit{LQ}) and unpolarized for the background ($t\bar{t}$)) is clearly seen here. 

There are few  interesting things to be noticed in this comparison of spectrum analysis for the LHC and the FCC collider. First, from the Table~\ref{Tab:Cut-efficiency flow table} we see that there is a significant increase of the pre-cut cross-section for \textit{LQ} in the FCC analysis compared to the LHC case, it is almost $18000$ times larger, whereas the same for the $t\bar{t}$ is almost $40$ times only.  Second, although the efficiency of the signal events remains the same for both \texttt{LHC} and \texttt{FCC}, that of the background increases largely for the \texttt{FCC} analysis. Third the final efficiency of getting the $b$-energy ratio for \textit{LQ}s are comparable in both cases. Fourth and most importantly, the plots show similar nature in both collider analysis and thus can be used to probe the chirality of the \textit{LQ}s in the collider analysis depending on the sensitivity.

\subsection{MET cut analysis}

We know that in collider, missing transverse energy comes from the particles which either do not or weakly interact with the electromagnetic and strong forces and thus escape the detection. Thus in collider physics it is an extremely important observable for discriminating leptonic decays of $W$ bosons and top quarks from background events such as multijet and Drell–Yan events, because they do not contain neutrinos. As we have considered specifically the hadronic decays with neutrino of our leptoquark, the MET analysis provide us very useful information about separating the signal events from the possible backgrounds. The basic idea is that, if there are neutrinos which are the decay product of the \textit{LQ}s, then they will contribute significantly towards generating a MET cut plot of significant slope and a clear signature of the presence of the leptoquarks as the parent particles.

\begin{table}[h!]
\centering
\resizebox{\textwidth}{!}
{\begin{tabular}{ |c|c|c|c|c| } 
 \hline
Collider &  \textit{LQ} Energy cut Eff & \textit{LQ} MET cut Eff & \textit{t$\bar{t}$} Energy cut Eff & \textit{t$\bar{t}$} MET cut Eff\\
 \hline 
 LHC $(\sqrt{s} = 13~\TeV)$ & 0.32 & 0.97 & 0.36 & 0.02 \\ 
 \hline
 FCC $(\sqrt{s} = 100~\TeV)$ & 0.36 & 0.98  & 0.34 & 0.02\\
 \hline
\end{tabular}}
\caption{Cut-efficiency flow table for signal and background after MET cut at  different colliders}
\label{Tab:MET Cut-efficiency flow table}
\end{table}

\begin{figure}[H]
\centering
\hspace*{\fill}
\begin{subfigure}{.45\textwidth}
  \centering
  \includegraphics[width=6cm]{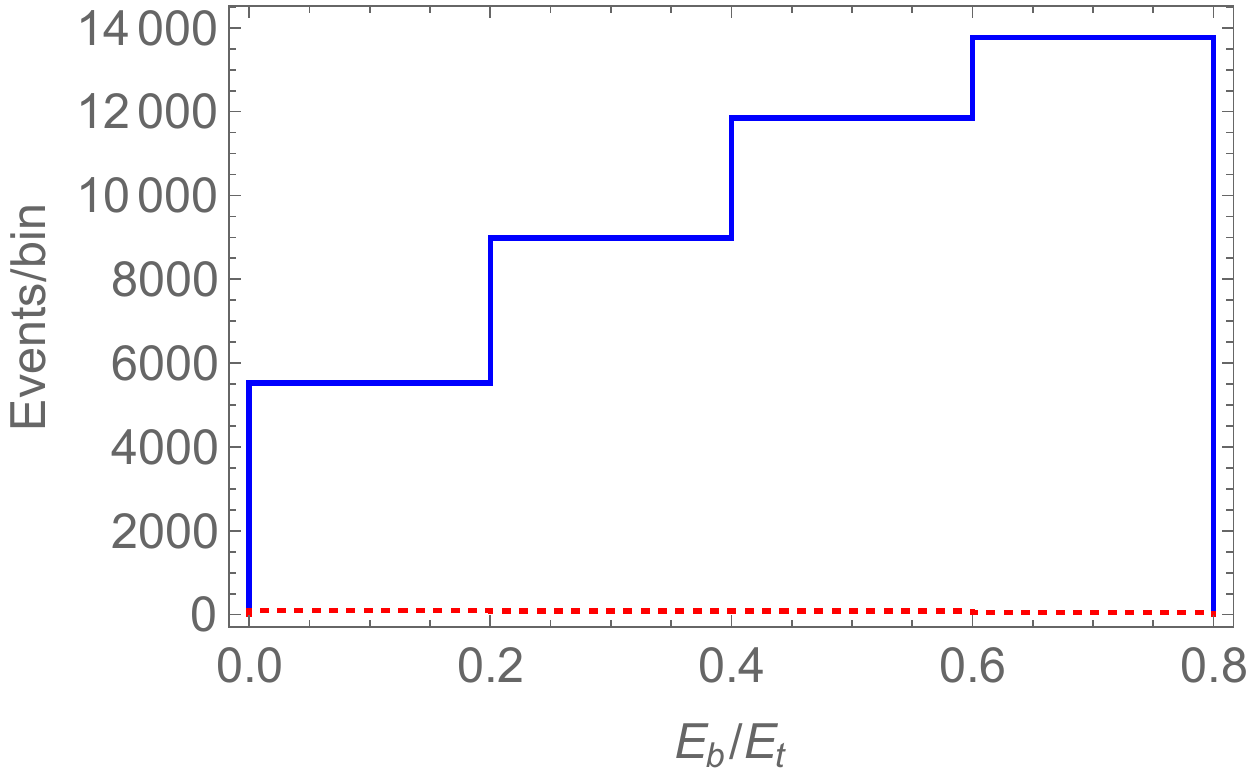}
  \caption{ MET cut $150 \GeV$ for LHC.}
  \label{fig:sub1}
\end{subfigure}%
\hfill
\begin{subfigure}{.45\textwidth}
  \centering
  \includegraphics[width=6cm]{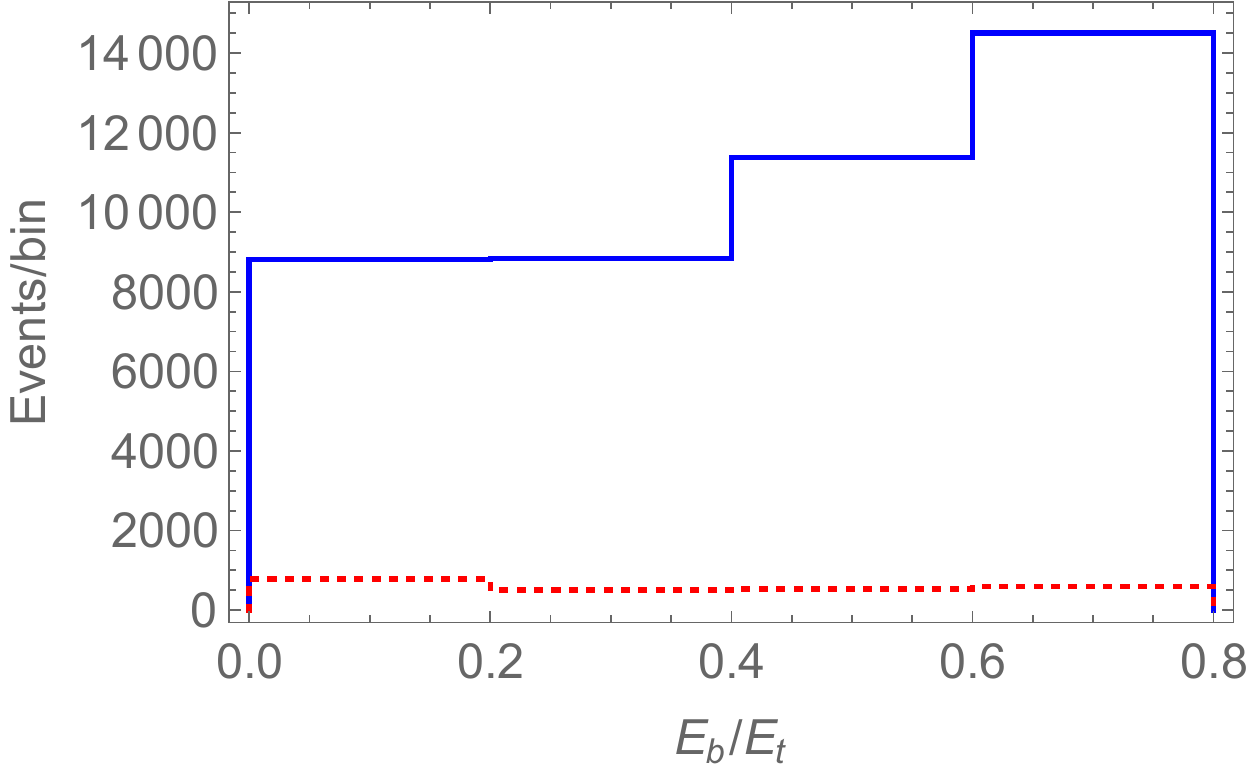}
  \caption{MET cut $150 \GeV$ for FCC.}
  \label{fig:sub2}
\end{subfigure}
\hspace*{\fill}
\caption{ Comparison of signal and background events with MET cut $150 \GeV$ at the LHC and FCC}
\label{fig:test}
\end{figure}

This clearly follows from the table as well as from the following figures. Table~\ref{Tab:MET Cut-efficiency flow table} shows although the energy cut efficiencies are comparable, the MET cut efficiency of the background is negligible compared to the signal. From Fig.~\ref{fig:sub1} also we find that for LHC analysis putting a MET cut of $150 \GeV$ the backgrounds can be completely eliminated. If we compare this with Fig.~\ref{fig:LHCnoMET} we see that a clear signature of left-handed \textit{LQ} is obtained. Similar trend can also be observed for the FCC analysis.  Since we have considered only the hadronic decay of the \textit{LQ}s we are ignoring the possible leptonic and semi-leptonic decays of the backgrounds, although it is quite possible that even we consider them they can not pass through the cuts applied and contribute significantly. It is interesting to notice that, although in FCC the background is higher than the LHC as we see from Fig~\ref{Fig:LHC-FCC spectrum}, with the MET cut analysis we can get distinguish the $LQ$ signal from the background.

\subsection{Sensitivity analysis}

The whole analysis described above gave us two important informations. First, the shape of the \textit{LQ} signal, with a definite chirality, curve compared to the dominant background and second, the efficiency with different kinematic cuts. We can use this cut-efficiency to find some numbers relevant for the collider study. We did two such estimations here. First is the estimation of number of events for different luminosities at different colliders and the second is the sensitivity test to find the luminosity required to get the $(\chi^2)$ value with $95\%$ C.L.  The results are given in Table~\ref{Tab:Sensitivity table}.

To get the expected number of events at the LHC, we just used $N_{evts} = \sigma \times \rm BR \times \rm Efficiency \times \rm Luminosity (\mathscr{L})$ where $\sigma$ is the only production cross-section of the signal events at different center of mass energies, ($13 \TeV $) for LHC and ($100 \TeV $) for FCC-hh. Thus choosing the recent luminosity of $(150 fb^{-1})$ used by the LHC, we find there is a possibility of getting only $1$ event whereas for the FCC-hh we can get a significantly higher number of signal events, as shown in the table, at the luminosity of $(1000 fb^{-1})$ which is about half of the predicted luminosity for the first phase $(2500 fb^{-1})$ of FCC-hh.

To perform a sensitivity test, means how much luminosity is required to get similar results as our analysis, for different colliders we wrote a \textit{chi-square $(\chi^2)$} function which is a function of luminosity only, see Appendix \ref{Chisq} for details. That function is fed with the values of 4-bin analysis we got after putting all the cuts including energy and MET cut as described in Sec.~\ref{Sec:Analysis}. Finally we solve for the required luminosity against the number for $95\%$ C.L with four variables, since we used four bins for our analysis. 

\begin{table}[h!]
\centering
\resizebox{\textwidth}{!}
{\begin{tabular}{ |c|c|c| } 
 \hline
Collider  & Number of Events $(N_{evts})$ & Luminosity ($fb^{-1})$ \\
 \hline 
 LHC $(\sqrt{s} = 13~\TeV)$ & $1.5 ~(150 fb^{-1})$ & $3.64 \times 10^5$ \\ 
 \hline
 FCC-hh $(\sqrt{s} = 100~\TeV)$ & $ 2.2\times 10^5 ~(1000 fb^{-1})$ & $0.363 $ \\ \hline
\end{tabular}}
\caption{Sensitivity analysis for different colliders}
\label{Tab:Sensitivity table}
\end{table}

%%%%%%%%%%%%%%%%%%%%%%%%%%%%%%%%%%%%%%%%%%%%%%%%%%%%%%%%%%%%

\section{Conclusion}
\label{Sec:conclusion}

Using a purely left-handed vector LQ which can only couple to the third generation of quark and lepton, we have shown that it can decay to a pure left-handed top quark which in turn decays to a more energetic bottom quark. Since top quarks decay before the hadronization, we can reconstruct its energy and form a model-independent variable, $b$-energy ratio which can be used as a clear distinguishable signature of tops chirality and consequently the chirality of the leptoquark of which top is the decay product.

In the FCC analysis the SM \textit{t$\bar{t}$} background is significantly higher but from the nature of the spectrum the \textit{LQ} signals are distinguishable. Following Table \ref{Tab:Cut-efficiency flow table} we see there is a significant increase of two-top system efficiency which may contribute to this high background for the FCC-hh case, though this might be a topic of future investigation. Further with the MET cut analysis we have seen that we can reduce the backgrounds significantly and have a clear signature of left-handed signature.   

Following Table~\ref{Tab:Sensitivity table} It is clear that the present or future LHC run are not sensitive to such analysis and the probability of getting sufficient signal events is almost negligible compared to the future collider like FCC which is quite sensitive for such production.

\section*{Acknowledgements}

I sincerely thank Yu Gao for putting forward the idea of this topic and providing some necessary tools to perform the analysis. I Also thank Lorenzo Calibbi and Ioannis Tsinikos for their time to go through the manuscript and providing some very  valuable comments and suggestions.

\appendix

\textbf{\LARGE{Appendix}}

%%%%%%%%%%%%%%%%%%%%%%%%%%%%%%%%%%%%%%%%%%%%%%%%%%%%%%%%%%%

\section{$\chi^2$ (\textit{chi-square}) Analysis} \label{Chisq}

The quantity known as \textit{chi-square} ($\chi^2$) is defined as 

\be \label{eq:chisquare}
\chi^2 \equiv \frac{(x_1 - \mu_1)^2}{\sigma_1^2} + \frac{(x_2 - \mu_2)^2}{\sigma_2^2} + \dots + \frac{(x_{\nu} - \mu_{\nu})^2}{\sigma_{\nu}^2}
\ee
where $\nu$ is the no of independent variables, $x_i$s are the independent variables, $\mu_i$s are the mean of each variable and $\sigma_i^2$ are the variance of each variable.

For our purpose the $x_i$s are the theoretical prediction which ideally should include both the production cross-section $\times \mathrm{BR}$ for the leptoquark production, known as signal, and the corresponding background. Assuming the background is eliminated we consider only the signal for $x_i$s. Therefore explicitly,

\be \label{xis}
x_i = \big(\sigma_{\rm Sig}+\sigma_{\rm Bkg}\big)\times (\mathscr{L})\times\textrm{Fraction of events} (N_i)
\ee
with 
$N_i = \frac{\textrm{Number of events in each bin}}{\textrm{Total number of events after MET cut}}$ and
\be
\sigma_{\rm Sig} = \sigma (p p > LQ \bar{LQ})\times \rm{BR}(LQ > t\nu\cdots)\times \rm{MET~Eff},
\ee
similar for the $\sigma_{\rm Bkg}$.

Similarly $\mu_i$s in this case are the SM background obtained from $t\bar{t}$ pair production and given by

\be \label{mus}
\mu_i = \sigma_{\rm Bkg}\times \textrm{Luminosity} (\mathscr{L})\times\textrm{Fraction of events} (N_i).
\ee

Therefore using equations \ref{xis} and \ref{mus} the final $\chi^2$ expression for our analysis of 4 bins become

\be \label{eq:chisqare4bins}
\chi^2_{4 \rm bins} = \sum_{i=1,4}\frac{(x_i - \mu_i)^2}{\sigma_i}
\ee
where $\sigma_i = x_i + \mu_i $.

\end{document}